\documentstyle[preprint,aps,amsmath,amsfonts,latexsym,amssymb,bbold,
graphicx,rotating,epsf]{revtex}

\begin{document}
\begin{center}
{\large\bf Tunneling of Born--Infeld Strings to $D2$--Branes}

{\it D.K. Park,$^{1,2}$\footnote{e-mail:
dkpark@hep.kyungnam.ac.kr}
S. Tamaryan,$^{1}$\footnote{e-mail:sayat@physik.uni-kl.de}
Y.--G. Miao$^{1}$\footnote{e-mail:miao@physik.uni-kl.de}
and
H. J. W. M\"{u}ller-Kirsten$^{1}$\footnote{e-mail:
mueller1@physik.uni-kl.de}}

{\it 1. Department of Physics,
 University of Kaiserslautern, D-67653 Kaiserslautern, Germany\\
2.Department of Physics, Kyungnam University, Masan, 631-701, Korea}
\end{center}

\vspace{0.5cm}

{\centerline{\bf Abstract}}
A Born--Infeld theory describing a $D2$--brane 
coupled to a 4--form RR field strength is considered, 
and the general solutions of the static and Euclidean time
equations are derived and discussed. 
The period of the bounce solutions is shown
to allow a consideration of tunneling and 
quantum--classical transitions in the sphaleron
region. The order of such transitions, depending
on the strength of the RR field strength, is
determined. A criterion is then derived to
confirm these findings.

\vspace{0.2cm}



\vspace{3cm}

\section{Introduction}

Recently it was argued \cite{emp} that a fundamental string can be viewed
as a collapsed brane, or more precisely as a $p$--brane ($p\geq 2$)
with ($p-1$) spatial directions of its worldvolume (e.g. with
spatial part ${\mathbf{R}}\times S^{p-1}$) collapsed to
zero size ($S^{p-1}$ shrunk to zero).  This view is somehow
a reversal of the traditional view of considering $D$--branes as
extended objects on which open
strings can put their ends \cite{pol}, and whose
low energy dynamics is described by
nonlinear Born--Infeld theory on
the worldvolume of the brane \cite{lei}. The latter
view implies, as has been shown \cite {callanmaldacena}
the existence of string--like configurations.
In ref. \cite{emp}
the reversal of the traditional view
was considered and the question was asked
whether by application of external forces (here an RR--spacetime
gauge potential) a string may expand
back into a $D$--brane, and this was -- in fact -- 
shown to be possible in the $D2$--brane case considered
via a tunneling process.

In the following our objective is to investigate
in more generality the problem posed in ref. \cite{emp}
and its solutions and the tunneling
process, and to determine
the order of the quantum--classical
transition (in analogy with phase transitions) at
nonzero temperatures as a function of the external
force.

The decay of a Born--Infeld brane--antibrane system
by tunneling was discussed earlier in ref.\cite{callanmaldacena}
but was
in that context discarded as being
 physically irrelevant.  The $D2$--brane model
of \cite{emp}
, however, seems a fascinating
and computationally manageable model
which allows the investigation
of various aspects of tunneling
related to strings.

In Section 2 we describe the formulation of the problem and its solutions. 
In Section 3 we calculate the wavelength and energy of the general static
configuration. In Section 4 we determine the
Euclidean version of the model for the torus brane
(defined by constant radius) and determine the
period of the bounce configuration. With this and
the sphaleron configuration 
we can determine the phase diagrams for
quantum classical transitions of the
string of vanishing radius into a brane of
constant radius.

It is found that a weak RR field leads to a
sharp transition 
described as of the first order, whereas
a stronger RR field leads to one
of the second order, described as
smooth. In Sections 5 and 6 
we then determine and discuss a criterion for the occurrence
of transitions of either order.
Finally in Section 7 we summarize
our findings in some conclusions.

\section{Formulation of the problem and its solutions}

Our starting point is the same Born--Infeld action describing
a $D2$--brane coupled to the RR--background spacetime 3--form gauge
potential $A_{\mu\nu\rho}$ with field strength $H=dA$ as that of ref.
\cite{emp}; we also use the same conventions (e.g.
$\alpha^{\prime}=1$).  Thus the action we consider is
\begin{equation}
 I = -\frac{1}{4{\pi}^2g}
\int d^3\xi
\bigg\{\sqrt{-det\bigg(g^{ind}_{\alpha\beta} +
2\pi{\cal F}_{\alpha\beta}\bigg)}
+\frac{1}{3!}\epsilon^{\alpha\beta\gamma}A_{\mu\nu\rho}\partial_{\alpha}
X^{\mu}\partial_{\beta}X^{\nu}\partial_{\gamma}X^{\rho}\bigg\}.
\label{1}
\end{equation}
Here $\mu,\nu,\rho = 0,\cdot\cdot\cdot,9$ are spacetime indices, and
$\alpha,\beta,\gamma =0,1,2$ worldvolume indices, $g$ is
the (type IIA) string
coupling and the dilaton is taken to be
constant.  The induced metric is the pullback of the 
spacetime metric (here assumed to be flat) and thus is
${\partial}_{\alpha}X^{\mu}{\partial}_{\beta}X_{\mu}$. The $U(1)$
gauge field tensor ${\cal F}_{\alpha\beta} $ contains in particular
the electric field ${\cal E}(t,z):=2\pi(\partial_0A_z-\partial_zA_0)$.
The background spacetime gauge field $H$ is taken to be
$H_{0123}=h$=const. As in ref.\cite{emp} we choose the world volume
to be cylindrical and hence define
$$
X^0=t,\;\; X^1=z,\;\; X^2=R(t,z)\cos\sigma,\;\; X^3=R(t,z)\sin\sigma,
$$
and all other $X^i$ = const., and $R>0$, $ 0<\sigma<2\pi$.  We also restrict
our considerations here only to those of
the electric field component of the
gauge field $A_{\alpha}$.  The action in terms of the
 worldvolume coordinates $t,z,\sigma$ with
$\sigma$ integrated out then becomes
\begin{equation}
I=\int dt \int dz {\cal L},\;\; {\cal L}= -\frac{1}{2\pi g}
\bigg(R\sqrt{1-{\dot R}^2-{\cal E}^2 + {R^{\prime}}^2} -\frac{h}{2}R^2\bigg)
\label{2}
\end{equation}
where dots and primes denote derivatives with respect to $t$ and $z$
respectively.  The Lagrangian density
${\cal L}(R, A_0,A_z, {\dot R}, {\dot A}_0, {\dot A}_z)$
yields in particular the Gauss law type of equation
\begin{equation}
\frac{\partial D}{\partial z} = 0,\;\; D = \frac{{\cal E}R}{\sqrt{
1-{\dot R}^2-{\cal E}^2 + {R^{\prime}}^2}} = const.
\label{3}
\end{equation}
In the static case to be considered below the constant
$D\equiv e$ can be looked at as a charge per radian, i.e.
$2\pi e\delta(z) A_0$
would be the appropriate source term in the
static Lagrangian.  Solving eq.(\ref{3}) for ${\cal E}$ one finds
\begin{equation}
{\cal E} = \frac{D\sqrt{1-{\dot R}^2 + {R^{\prime}}^2}}{\sqrt{D^2+R^2}}.
\label{4}
\end{equation}
Since $D$ is constant and so unaffected by the form of $R$ it is
convenient to express the action in terms of $D$ and regard the result as
a functional of $R$ so that the variation of the new
Lagrangian density ${\tilde {\cal L}}$ with respect to ${\cal E}$
is zero.  The appropriate Legendre transform of ${\cal L}$ is
\begin{equation}
{\tilde {\cal L}}= {\cal L} -\frac{D{\cal E}}{2\pi g}
\label{5}
\end{equation}
since with eq.(\ref{3}) $\partial{\cal L}/{\partial 
\cal E} = \frac{D}{2\pi g}$.  Evaluating ${\tilde {\cal L}}$
with the help of eq. (\ref{4}) one obtains as in \cite{emp}
\begin{equation} 
{\tilde {\cal L}}= - \frac{1}{2\pi g}\bigg(\sqrt{R^2+D^2}
\sqrt{1-{\dot R}^2 + {R^{\prime}}^2}-\frac{h}{2}R^2\bigg).
\label{6}
\end{equation}
Since ${\cal L}$ is a functional of $R, A_0, A_z$, we have
conjugate momenta $P=\partial{\cal L}/\partial{\dot R},
P_{A_0}=\partial{\cal L}/\partial{{\dot A}_0} =0$
and
$P_{A_z}=D/g$. As usual the vanishing of $P_{A_0}$ implies the
Gauss law as the constraint (\ref{3}), so that the 
Hamiltonian $H$ is defined by
\begin{equation}
H=\int dz\bigg(P{\dot R}+P_{A_z}{\dot A}_z-{\cal L}
+\frac{1}{g}A_0\frac{\partial D}{\partial z}\bigg)
\label{7}
\end{equation}
where $A_0$ acts as Lagrange multiplier.
Evaluating this, one obtains
\begin{equation}
H=\frac{1}{2\pi g}\int dz\bigg \{\frac{(1+{R^{\prime}}^2)\sqrt{D^2+R^2}}
{\sqrt{1-{\dot R}^2 + {R^{\prime}}^2}}-\frac{h}{2}R^2\bigg\}.
\label{8}
\end{equation}
The static energy $E$ is therefore
\begin{equation}
E=\frac{1}{2\pi g}\int dz \bigg\{\sqrt{(1+{R^{\prime}}^2)(D^2+R^2)}
-\frac{h}{2}R^2\bigg\}.
\label{9}
\end{equation}
Variation of $E$ with respect to $R$ yields the equation for the
static configuration
\begin{equation}
\sqrt{\frac{R^2+D^2}{1+{R^{\prime}}^2}}-\frac{h}{2}R^2 = C
\label{10}
\end{equation}
where $C$ is an integration constant.  We rewrite this equation
\begin{equation}
R^{\prime}=\stackrel{-}{(+)}\frac{h}{hR^2+2C}\sqrt{(R^2_+-R^2)(R^2-R^2_-)},
\;\;R^{\prime}\neq 0
\label{11}
\end{equation}
with
\begin{equation}
{R_{\pm}}^2=\frac{2}{h^2}\bigg[(1-Ch)\pm\sqrt{1-2Ch+h^2D^2}\bigg]
\label{12}
\end{equation}
and assume for simplicity
that $C\geq D$.
We note also for later reference
\begin{equation}
R^2_+R^2_-=\frac{4}{h^2}(C^2-D^2),\;\; R^2_++R^2_-=\frac{4}{h^2}(1-Ch).
\label{13}
\end{equation}
Many special solutions have already been discussed in ref.\cite{emp}.
E.g. for $h=0$ the case $C=D$ implies spikes $R=R_0\exp ({\pm  z/C})$
for $z<0 $ or $z>0$ 
respectively corresponding to
 the BIon in $D3$ Born--Infeld theory\cite{park}, and the case
$C^2>D^2$ implies catenoids
$R=\sqrt{C^2-D^2}\cosh (z-z_0)$.
Here we are interested in the general solution. However, we note
the possibility of the two signs of $R^{\prime}$ in eq.(\ref{11}).
The significance of these is the same as those of the
catenoid discussed in\cite{park}. There one sign gives the spike--like
solution which minimizes the action, the other sign the 
solution which maximizes the action.  The socalled wormhole solution is
the two--spike solution obtained by matching  the two at a
suitable point.

We now consider the general solution of
the static equation (\ref{10}).  Assuming $R_-\leq R\leq R_+$
and $R(z_0)=R_+, $ eq.(\ref{10}) becomes
\begin{equation}
\int^{R_+}_R dR\frac{hR^2+2C}{\sqrt{(R^2_+-R^2)(R^2-R^2_-)}}=-h(z_0-z).
\label{14}
\end{equation}
We observe that for real values of $z$ the solution has 
turning points at $R = R_+$ and $R_-$. The distance between
these therefore defines half a wavelength $\lambda$ of
oscillations between these.

The left hand side of eq. (\ref{14}) can
 be evaluated in terms of elliptic integrals.
Then the general solution is 
given by (cf. ref.\cite{elliptic}, pp.56, 300-301)
\begin{equation}
hR_+E(\psi, k) + \frac{2C}{R_+}F(\psi,k) = - h(z_0-z).
\label{15}
\end{equation}
Here $F(\psi, k)$ and $E(\psi, k)$ are the incomplete elliptic integrals
of the first and second kinds respectively, $k$ is the
elliptic  
modulus given by
$$
k=\frac{\sqrt{R^2_+-R^2_-}}{R_+}
$$
and $\psi$ is the angle defined by
$$
\psi = \sin^{-1}\sqrt{\frac{R^2_+-R^2}{R^2_+-R^2_-}}.
$$
We now consider two important limiting cases:
a) The limit C=D. In this case
$$
R_+=\frac{2}{h}\sqrt{1-Dh},\;\; R_-=0,\;\; k=1
$$
and
$$
\sin\psi=\sqrt{R^2_+-R^2}/{R_+}.
$$
Using $E(\psi, 1) = \sin\psi, F(\psi, 1) =\ln (\tan\psi + \sec\psi)$
eq.(\ref{15}) becomes
\begin{equation}
\sqrt{R^2_+-R^2}+\frac{2D}{hR_+}\ln\frac{R_++\sqrt{R^2_+-R^2}}{R} = -(z_0-z)
\label{16}
\end{equation}
which agrees with a result of ref.\cite{emp}
and describes a BI--string with spheroidal
bulge centered at $z_0$
where $R=R_+$.
For $R\rightarrow R_-=0$ the solution becomes
string--like (this case may be dubbed `vacuum solution').
b) The limit $C=(1+h^2D^2)/2h $ is the limit
of the sphaleron configuration,
i.e. that with zero separation between the
turning points.  In this case
$$
R_-=R_+=\frac{\sqrt{2(1-Ch)}}{h},\;\; k=0.
$$
Using
$$
E(\psi, 0) = F(\psi, 0) = \psi
$$
eq.(\ref{15}) becomes
$$
\bigg(hR_++\frac{2C}{R_+}\bigg)\psi = -h(z_0-z),
\;\; i.e.\; \frac{2}{\sqrt{1-h^2D^2}}\psi = -h(z_0-z)
$$
which implies
$$
R^2 = R^2_+-(R^2_+-R^2_-)\sin^2\frac{h\sqrt{1-h^2D^2}(z-z_0)}{2}=R^2_+
$$
so that
$$
R=R_+=\frac{\sqrt{2(1-Ch)}}{h}=\frac{1}{h}\sqrt{1-h^2D^2} \equiv R_S
$$
where $R_S$ is the sphaleron configuration, 
in agreement with a result discussed in \cite{emp} and of particular
interest for $h\neq 0$.

\section{Calculation of wavelength and energy
 of the general static configuration}

Having prepared the ground for our considerations we now proceed to
calculate the wavelength $\lambda $ 
of the general static solution of eq.(\ref{14}). 
As mentioned above, the points $R_+, R_-$ are turning points,
so that with $R$ in eq.(\ref{14}) replaced by
$R_-$, we can put $z-z_0 = \lambda/2$.  Then 
\begin{eqnarray}
\int^{R_+}_{R_-} dR\frac{hR^2+2C}{\sqrt{(R^2_+-R^2)(R^2-R^2_-)}}
&=& hR_+E(\frac{\pi}{2}, k) +2C\frac{1}{R_+}F(\frac{\pi}{2}, k)\nonumber\\
&=&hR_+E(k)+\frac{2C}{R_+}K(k) =h\frac{\lambda}{2}
\label{17}
\end{eqnarray}
where $E(k)$ and $K(k)$ are the complete elliptic integrals of the
second and first kinds respectively. The wavelength $\lambda$
can therefore be expressed as
\begin{equation}
\lambda = 2R_+E(k) + \frac{4C}{hR_+} K(k).
\label{18}
\end{equation}
Using the special values $E(1)=1, K(1)=\infty, E(0)=K(0)=\pi /2$
(cf. \cite{elliptic}, p.10), we can obtain the values of $\lambda $
of the two limiting cases discussed above. Thus for a) with $C=D$, one
obtains $\lambda = \infty$ or frequency $\Omega = 0$, and for b) with
$C=(1+h^2D^2)/(2h)$, the sphaleron limit, one obtains
$\lambda = 2\pi/(h\sqrt{1-h^2D^2})$. With this we can define
as sphaleron frequency
$$
\Omega_{sph}=h\sqrt{1-h^2D^2}.
$$

Next we compute the energy of the solution (\ref{14})
by replacing ${R^{\prime}}^2$ in $E$ by the expression 
obtained from  eq.(\ref{10}),
i.e.
\begin{equation}
{R^{\prime}}^2 = \frac{4(R^2+D^2)}{(hR^2+2C)^2} - 1.
\label{19}
\end{equation}
We rewrite the expression (\ref{9}) with the contribution
of the fundamental BI string
(the limit $C-D = 0, R=R^{\prime}=0$) separated in view of its
divergence
in the linear case, i.e. we write 
with a somewhat arbitrary subdivision
\begin{equation}
E = \frac{D}{2\pi g} \int dz + U_0,\;\;  U_0 = \frac{1}{2\pi g}\int dz
\bigg[\frac{2(R^2+D^2)}{hR^2+2C}-\frac{h}{2}R^2 - D\bigg].  
\label{20}
\end{equation}
Here the first term in $E$ divided by the length
$\int dz$ gives the tension of the BI string, i.e.
$T_{BI} = D/2\pi g$. As shown in ref.\cite{callanmaldacena}
the quantization condition on the $D$--flux is
$$
\frac{1}{2\pi}\int_{S^1} D = gn, \;\; i.e. \;\; D=gn,
$$
where $n$ is the number of fundamental strings,
each of tension $1/2\pi\alpha^{\prime}$, adsorbed by the
BI string.  Thus the first
contribution to $E$ in eq.(\ref{20}) is the energy
of $n$ fundamental strings. 
Then by explicit calculation and
 using $\int^{\lambda}_0dz = -2\int^{R_+}_{R_-}dR/R^{\prime}$
and again replacing $R^{\prime}$ by the expression of eq. (\ref{19}),
we obtain (cf. Appendix A)
\begin{eqnarray}
U_0=\frac{1}{6\pi gh}\bigg[&&\frac{12D(D-C)+(1-k^2)h^2R^4_+}
{R_+} K(k)\nonumber\\
&+& 2R_+\bigg\{3\bigg[2-h(C+D)\bigg]-(2-k^2)h^2R^2_+\bigg\} E(k)\bigg].
\label{21}
\end{eqnarray}
Of course, the $R^2_-$--dependence is still contained
in $k^2$.  Again we consider the two limiting cases of above.
In case a), the limit $C=D$ with
$$\lim_{k\rightarrow 1} (1-k^2)K(k) = 0,$$
one can show directly that
\begin{equation}
U_0=\frac{h}{6\pi g}R^3_+,\;\; R_+=\frac{2}{h}\sqrt{1-Dh},\;\; k=1,
\label{22}
\end{equation}
which is consistent with eq.(\ref{15}) of ref.\cite{emp}
and represents a bulk energy.
In case b), the sphaleron limit $C=(1+h^2D^2)/2h$, direct calculation
yields
\begin{equation}
U_0=\frac{(1-hD)^2}{2R_+gh^3},\;\; R_+=\frac{\sqrt{2(1-Ch)}}{h}
=\frac{\sqrt{1-h^2D^2}}{h}, \;\; k=0
\label{23}
\end{equation}
To show that this is physically relevant, we insert
the constant value
$R=R_S=\sqrt{1-h^2D^2}/h$ into the expression for the static energy
$E$ of eq.(\ref{9}) and
integrate over a length $L$. Then $E$ becomes
$$
E=\frac{1+h^2D^2}{4\pi gh}L.
$$
From eq. (\ref{20}) we know that
this energy has to be
$$
E=\frac{D L}{2\pi g} + U_0,
$$
i.e. this has to be satisfied.  Equating the two expressions we
obtain
$$
U_0=\frac{(1-hD)^2}{4\pi gh}L.
$$
If we put here $L=\lambda = 2\pi/{h\sqrt{1-h^2D^2}}$, we
obtain
$$
U_0=\frac{(1-hD)^2}{2R_+gh^3}
$$
in agreement with eq.(\ref{23}).

\section{The quantum--classical decay rate transition
to the torus brane}
 
In the following we consider first the
solutions to the Euclidean version of
the action (\ref{1}), then specialise to the torus brane defined by the
condition $R^{\prime}=0$, obtain the corresponding
periodic bounce solutions, and then the phase diagrams. 

We are interested in solutions to the Euclidean version
of the action (\ref{1}), i.e.
(cf. eq. (\ref{6}) )
\begin{equation}
{\tilde I}_E(D,R)=\frac{1}{2\pi g}
\int d\tau dz\left[\sqrt{(R^2+D^2)(1+{\dot R}^2+{R^{\prime}}^2)}-
\frac{h}{2}R^2\right]
\label{24}
\end{equation}
where ${\dot R}$ now denotes differentiation with respect to Euclidean time
$\tau =-it$.  The equation of motion derived from
(\ref{24}) is
\begin{eqnarray}
\frac{R(1+{\dot R}^2+{R^{\prime}}^2)}
{\sqrt{(R^2+D^2)(1+{\dot R}^2+{R^{\prime}}^2)}}
&-&\frac{\partial}{\partial\tau}\frac{{\dot R}(R^2+D^2)}
{\sqrt{(R^2+D^2)(1+{\dot R}^2+{R^{\prime}}^2)}}\nonumber\\
&-&\frac{\partial}{\partial z}
\frac{R^{\prime}(R^2+D^2)}
{\sqrt{(R^2+D^2)(1+{\dot R}^2+{R^{\prime}}^2)}}
-hR = 0.
\label{25}
\end{eqnarray}
Here we consider only tunneling to the torus brane ($R^{\prime} = 0$).
In this case the equation reduces to
\begin{equation}
\frac{R(1+{\dot R}^2)}
{\sqrt{(R^2+D^2)(1+{\dot R}^2)}}
-\frac{\partial}{\partial\tau}\frac{{\dot R}(R^2+D^2)}
{\sqrt{(R^2+D^2)(1+{\dot R}^2)}}
-hR =0.
\label{26}
\end{equation}
After some manipulation one can show that eq. (\ref{26})
is exactly the same as eq. (\ref{11}) if $z$ there is replaced by $\tau$.
In fact this is easily understood from the symmmetry
of the Euclidean action (\ref{24}) under the exchange
$\tau \leftrightarrow z$.
Hence the periodic bounce solution is  in this case
obtained directly from eq.(\ref{15}) by changing $z_0-z$ into $\tau_0-\tau$,
i.e. from
\begin{equation}
hR_+E(\psi,k) + \frac{2C}{R_+} F(\psi,k) = - h(\tau_0-\tau).
\label{27}
\end{equation}
With this expression for Euclidean time $\tau_0-\tau$ we
can calculate the period P.
We first consider the Euclidean action for this configuration
with $R^{\prime} =0$ (as stated earlier) which is
\begin{eqnarray}
{\tilde I}_{E,cl}(D,R_{cl})
&=&\frac{1}{2\pi g}\int d\tau dz
\left[\sqrt{{(R_{cl}}^2+D^2)
(1+{\dot R_{cl}}^2)}-\frac{h}{2}{R_{cl}}^2\right]\nonumber\\
&=&L\left[\frac{D}{2\pi g}\int dz + U_0\right]
\label{28}
\end{eqnarray}
where $U_0$ is given by eq. (\ref{21}) and $L$ is
now  the circumference of the
torus in Euclidean time.
  We write the part with the contribution of fundamental
strings subtracted out
\begin{equation}
{\tilde I}^{(0)}_{E,cl}(D,R_{cl}) = LU_0(C,D,h).
\label{29}
\end{equation}
As in statistical mechanics the period $P$ of this configuration is
to be identified with the reciprocal of temperature $T$
(cf. ref.\cite{gor}).  
Since the wavelength 
$\lambda= 2\pi/\Omega$, where $\Omega$ is the corresponding 
angular frequency, we can rewrite eq. (\ref{27}) here 
\begin{equation}
P = 2R_+E(k) + \frac{4C}{hR_+}K(k) = \frac{1}{T}.
\label{30}
\end{equation}
The sphaleron configuration in this tunneling is $$
R = R_S = \frac{\sqrt{1-h^2D^2}}{h}.
$$
The barrier height or sphaleron energy is
\begin{equation}
E_{sph} = \frac{L}{4\pi gh}(1-hD)^2
\label{31}
\end{equation} 
as observed at the end of Section 3.
From this we obtain for the sphaleron action
\begin{equation}
I^{(0)}_{sph} = \frac{E_{sph}}{T}.
\label{32}
\end{equation}
With the formulas at hand we can now plot the phase diagrams.The integration
constant $C$ here plays the same role as the energy $E$ of a pseudoparticle
in the usual quantum mechanics around such
a configuration
 ($E = 0$ in the latter being its value for the vacuum solution, and
$E = E_{sph}$ that for the sphaleron, see e.g. \cite{mott}
and \cite{supermini}).  
In the present case the vacuum solution has $C = D$, and the
sphaleron $C= (1+h^2D^2)/2h$.
In Fig.1(a)  we show the behaviour of the period $P$
as a function of $C$ for $D=1$ and $h=0.7$. We observe
a monotonically decreasing behaviour characteristic of
a smooth second order transition. In Fig. 1(b) we
plot for the same values of $D$ and $h$
the behaviour of the sphaleron action per unit length   
$I^{(0)}_{sph}/L$ and the Euclidean action per unit length
${\tilde I}^{(0)}_{E,cl}/L$ versus temperature $T$.
We observe the expected smooth transition from one to the
other.
In Figs.2(a) and (b) we plot the same quantities
again for $D=1$ but for the smaller value $h=0.3$ of the 
RR field.  This time we observe a rise of the
period beyond a critical value, and a corresponding
nonsmooth bifurcation of the action implying
a sharp quantum classical transition, generally described
as of first order.

\section{The criterion for a sharp transition}

A criterion for the occurrence of sharp or first order quantum--classical
transitions was developed in ref. \cite{gor}. A more explicit
criterion useful for practical purposes was obtained in ref. \cite{rana}.
Our considerations here are based on this latter reference
where a detailed explanation of steps can be found.
Further applications of the criterion have been given
in refs. \cite{mott} and \cite{supermini}.
The criterion for occurrence of a first order transition
is $\Omega^2-\Omega^2_S > 0,$
where $\Omega_S$ is the sphaleron frequency and $\Omega $
the possible frequency different from $\Omega_S$ \cite{rana}.
 
With some manipulations eq.(\ref{26}) can
be put into the following form
\begin{equation}
{\ddot R}-R\frac{1+{\dot R}^2}{R^2+D^2}+hR(1+{\dot R}^2)
\sqrt{\frac{1+{\dot R}^2}{R^2+D^2}}=0.
\label{33}
\end{equation}
We expand this equation around the sphaleron
configuration $R_S$ by setting
\begin{equation}
R = R_S +\delta R(\tau), \;\;\; R_S=\frac{1}{h}\sqrt{1-h^2D^2}.
\label{34}
\end{equation}
Using the following expansions
\begin{eqnarray}
\frac{1}{R^2+D^2}&=&h^2\bigg[1-2h\sqrt{1-h^2D^2}
\delta R+h^2(3-4h^2D^2)\delta R^2\nonumber\\
&&+4h^3(2h^2D^2-1)\sqrt{1-h^2D^2}\delta R^3 +\cdot\cdot\cdot\bigg],\nonumber\\
\frac{1+{\dot R}^2}{R^2+D^2}&=&h^2\bigg[1-2h\sqrt{1-h^2D^2}\delta R\nonumber\\
&&+\bigg\{h^2(3-4h^2D^2)\delta R^2 +\delta{\dot R}^2\bigg\}\nonumber\\
&&+\bigg\{4h^3(2h^2D^2-1)\sqrt{1-h^2D^2}\delta R^3 
-2h\sqrt{1-h^2D^2}\delta {\dot R}^2\delta R\bigg\}\nonumber\\
&&+\cdot\cdot\cdot\bigg],\nonumber\\
\sqrt{\frac{1+{\dot R}^2}{R^2+D^2}}
&=&h\bigg[1-h\sqrt{1-h^2D^2}\delta R\nonumber\\
&&+\frac{1}{2}\bigg\{h^2(2-3h^2D^2)\delta R^2
+\delta {\dot R}^2\bigg\}\nonumber\\
&&-\frac{h}{2}\sqrt{1-h^2D^2}\bigg\{
(2-5h^2D^2)h^2\delta R^3+\delta R \delta {\dot R}^2\bigg\}+\cdot\cdot\cdot\bigg]
\label{35}
\end{eqnarray}
one then obtains
\begin{eqnarray}
R\frac{1+{\dot R}^2}{R^2+D^2}
=h\bigg[&&\sqrt{1-h^2D^2}-h(1-2h^2D^2)\delta R\nonumber\\
&+&\sqrt{1-h^2D^2}\bigg\{h^2(1-4h^2D^2)\delta R^2+\delta{\dot R}^2\bigg\}\nonumber\\
&-&h\bigg\{h^2(1-8h^2D^2+8h^4D^4)\delta R^3
+(1-2h^2D^2)\delta{\dot R}^2\delta R\bigg\}
\nonumber\\
&+&\cdot\cdot\cdot\bigg]
\label{36}
\end{eqnarray}
and
\begin{eqnarray}
R(1+{\dot R}^2)\sqrt{\frac{1+{\dot R}^2}{R^2+D^2}}
=&&\sqrt{1-h^2D^2}+h^3D^2\delta R\nonumber\\
&+&\sqrt{1-h^2D^2}\bigg\{-\frac{3}{2}h^4D^2\delta R^2 +\frac{3}{2}
\delta{\dot R}^2\bigg\}\nonumber\\
&+&\bigg\{h^3(2h^2D^2-\frac{5}{2}h^4D^4)\delta R^3
+\frac{3}{2}h^3D^2\delta R \delta{\dot R}^2\bigg\}\nonumber\\
&+&\cdot\cdot\cdot
\label{37}
\end{eqnarray}               
Eq. (\ref{33}) can now be expanded as
\begin{equation}
\delta {\ddot R} + h^2(1-h^2D^2) \delta R 
+ G_2(\delta R) + G_3(\delta R) + \cdot\cdot\cdot = 0
\label{38}
\end{equation}
where
\begin{eqnarray}
G_2(\delta R)&=&\frac{h\sqrt{1-h^2D^2}}{2}
\bigg[-h^2(2-5h^2D^2)\delta R^2 +\delta{\dot R}^2\bigg],\nonumber\\
G_3(\delta R)&=&\frac{h^2}{2}\bigg[h^2(2-12h^2D^2+11h^4D^4)\delta R^3
+(2-h^2D^2)\delta{\dot R}^2\delta R\bigg].
\label{39}
\end{eqnarray}
In lowest or linear order we have therefore
\begin{equation}
\delta {\ddot R} + h^2(1-h^2D^2) \delta R =0
\label{40}
\end{equation}
with solution
\begin{equation}
\delta R = a \cos\Omega_S\tau, \;\;\Omega_S=h\sqrt{1-h^2D^2}
\label{41}
\end{equation}
where $a$ is a small amplitude of oscillation around sphaleron.
We observe that the sphaleron period resulting from eq. (\ref{41}),
i.e.
$$
P_{sph}=\frac{2\pi}{\Omega_S} =\frac{2\pi}{h\sqrt{1-h^2D^2}}
$$
coincides exactly, as expected,
 with the wavelength defined after eq.(\ref{18}).

We proceed to the second order calculation by setting
\begin{equation}
\delta R = a\cos\Omega\tau + a^2\eta_1(\tau),\;\;
\Omega^2 = \Omega^2_S+a\triangle_1\Omega^2.
\label{42}
\end{equation}
with a new or as yet undetermined frequency $\Omega$.
Inserting this trial solution into eq.(\ref{38}) and reexpressing
squares of $\cos\Omega\tau$ by $\cos2\Omega\tau$ we obtain at
this level of the approximation
\begin{eqnarray} 
\eta_1=\bigg(\frac{\partial^2}{\partial\tau^2}+\Omega^2_S\bigg)^{-1}
\bigg[\triangle_1\Omega^2\cos\Omega_S\tau\nonumber\\
+\frac{\Omega_S}{4}\bigg\{[h^2(2-5h^2D^2)-\Omega^2]
+[h^2(2-5h^2D^2)&+&\Omega^2]\cos2\Omega\tau\bigg\}\bigg].
\label{43}
\end{eqnarray}
In order to avoid infinity and have a defined fluctuation,
 the square bracket on the right
hand side must not contain the zero mode of the operator
$\bigg(\frac{\partial^2}{\partial\tau^2}+\Omega^2_S\bigg)$.
Hence we must demand $\triangle_1\Omega^2=0$.  This condition therefore
yields
\begin{equation}
\Omega = \Omega_S
\label{44}
\end{equation}
to this order of the calculation
together with the fluctuation
\begin{equation}
\eta_1= g_1+g_2\cos 2\Omega_S\tau
\label{45}
\end{equation}
where
\begin{equation}
g_1=\frac{\Omega_S}{4}\frac{1-4h^2D^2}{1-h^2D^2},\;\;g_2=-\frac{\Omega_S}
{4}\frac{1-2h^2D^2}{1-h^2D^2}.
\label{46}
\end{equation}

We proceed to the third order by setting
\begin{equation}
\delta R=a\cos\Omega\tau + a^2\eta_1(\tau) +a^3\eta_2(\tau), \;\;
\Omega^2=\Omega^2_S+a\triangle_1\Omega^2+a^2\triangle_2\Omega^2.
\label{47}
\end{equation}
Inserting this ansatz into eq.(\ref{38}) yields with
a procedure as before but now at this level
\begin{equation}
\eta_2=\frac{1}{a^3}\bigg(\frac{\partial^2}{\partial\tau^2}+\Omega^2_S
\bigg)^{-1}\bigg[\chi^{(0)}_2+\chi^{(1)}_2\cos\Omega_S\tau + 
\chi^{(2)}_2\cos2\Omega_S\tau + 
\chi^{(3)}_3\cos3\Omega_S\tau\bigg],
\label{48}
\end{equation}
where (only this is needed here)
\begin{eqnarray}
\chi^{(1)}_2=a(\Omega^2-\Omega^2_S)
-a^3\bigg[&-&h^2(2-5h^2D^2)\Omega_S(g_1+\frac{g_2}{2})+\Omega^3_Sg_2\nonumber\\
&+&\frac{3}{8}h^4(2-12h^2D^2+11h^4D^4)+\frac{h^2}{8}\Omega^2_S(2-h^2D^2)\bigg]
\nonumber\\
=a^3\triangle_2\Omega^2 &-& a^3\bigg[1-4h^2D^2\bigg]\frac{h^4}{2}.
\label{49}
\end{eqnarray}
Again we set this equal to zero to avoid infinity in (\ref{48})
and so determine $\triangle_2\Omega^2$.  The criterion for a
sharp or first
order transition $\Omega^2-\Omega^2_S > 0$ then implies
\begin{equation}
\triangle_2\Omega^2 > 0, \;\; i.e. \;\;\;  hD < \frac{1}{2}.
\label{50}
\end{equation}
We observe that this result agrees with our earlier findings that
the smooth second order transition occurs for larger
values of the applied force or $h$. Thus also the bifurcation
point in the plot of the action must disappear in the large $h$
domain.  We now investigate this point in more detail.

\section{Tunneling of the string to an arbitrary $D2$--brane}

It is much more difficult to understand the tunneling associated with the
sphaleron of the general static solution (\ref{15}), i.e.
for general values of $\psi $ in
\begin{equation}
hR_+E(\psi, k) + \frac{2C}{R_+}F(\psi,k) = - h(z_0-z)
\label{51}
\end{equation}
However, we can understand  qualitatively this tunneling using
the tunneling to the torus investigated in the previous
sections. We first compute the barrier height which is
equal to the sphaleron energy, i. e.
\begin{eqnarray}
E_{sph}&=&\frac{1}{2\pi g}\int dz\bigg[\sqrt{(R^2+D^2)(1+{R^{\prime}}^2)}
-\frac{h}{2}R^2\bigg] - c.f.s.\nonumber\\
&=&\frac{1}{6\pi gh}\bigg[\frac{12D(D-C)+(1-k^2)h^2R^4_+}{R_+}K(k)\nonumber\\
&&+2R_+\bigg\{3[2-h(C+D)]-(2-k^2)h^2R^2_+\bigg\}E(k)\bigg]
\label{52}
\end{eqnarray}
where c.f.s. is the contribution of the
fundamental strings.
In fact, the expression (\ref{52})
 is exactly the same as the classical action for
tunneling to the torus divided by $L$, the circumference of the torus,
as one can see by comparison with eq. (\ref{21}). 
With this observation we can deduce the following facts.

1) In the region of $hD>\frac{1}{2}$ the energy $E_{sph}$ is minimal for
tunneling to the torus ($R^{\prime}=0$). Hence tunneling to the
torus may be the dominant process.

2) In the region of $hD<\frac{1}{2}$ the energy $E_{sph}$ 
has a minimum at the tunneling which corresponds to the
bifurcation point in the action--versus--temperature
diagram (cf. Fig. 2(b)). Thus in this region the
dominant tunneling may not be tunneling to the torus
but to a wiggly distorted brane.  

It is important therefore to understand when the 
bifurcation point appears. We can study this, for instance,
by determining those values of $C=C(D,h)$,
which correspond to the bifurcation points in
tunneling with $R^{\prime}=0$. In order to determine
this parameter $C$ as a function of $D$ and $h$, we
recall that the period $P$ of the bounce configuration
for a first order transition has an extremum period
as can also be seen from Fig.2(a) (or ref.\cite{prl}).  Thus this $C(D,h)$
is determined by
\begin{equation}
{\frac{dP}{dC}}\bigg|_{C=C(D,h)} = 0.
\label{53}
\end{equation}
Here $P$ is, of course, the same as the wavelength $\lambda$
of eq. (\ref{18}).
In Fig.3(a) the solution $C(D,h)$ of this equation is
plotted for $D=1$ as a function of $h$.  One can see that
the maximum of $C$ meets the bifurcation point
at $C=1.2$ (cf. Fig.2(a)) at exactly $h=0.5$, the
limit set by the criterion eq. (\ref{50}). Fig. 3(b)
shows a corresponding plot for $D=2$.  Both of these
plots therefore show how the bifurcation point
disappears in the large $h$ region, i.e. beyond
the limit given by the criterion (\ref{50}).
Finally, in Fig. 4  we show the
shape of a brane at a bifurcation point.

\section{Conclusions}
The expansion of a string into a $D2$--brane by tunneling
through a barrier provided by a 4--form RR field strength
was already shown to be possible in ref.\cite{emp}.  In the above
we have investigated this problem in more detail by 
deriving the explicit solutions of the classical equation in
terms of elliptic functions. These solutions then
permitted the derivation of the period of the
bounce configurations, as well as a detailed
study of the sphaleron at the top of the   
potential barrier. With these tools it was
possible to study the quantum--classical behaviour
at nonzero temperatures 
in the sphaleron domain, and to determine the
order of the transitions (in analogy with
phase transitions) depending on the magnitude
of the applied RR field.  We then confirmed
these findings with the derivation of a criterion
for the occurrence of such transitions in the
present case.
The case of the $D2$--brane
discussed here has the advantage of permitting explicit
calculations with relative ease. We expect the
method, however, to be applicable also to more
complicated cases.

\vspace{4cm}

{\bf Acknowledgements:} D.K. P. and S.T. acknowledge support by the
Deutsche Forschungsgemeinschaft (DFG), Y.--G. M acknowledges
support by an A. von Humboldt research fellowship. D.K. P. is also
supported by Korea Research Foundation Grant (KRF-2000-D00073).

\newpage

\begin{appendix}{\centerline{\bf Appendix A}}
\setcounter{equation}{0}
\renewcommand{\theequation}{A.\arabic{equation}}

Here we indicate briefly how eq.(\ref{21}) is obtained.
Substituting $R^{\prime}$ of eq.(\ref{11})
into the expression
(\ref{20}) for $U_0$ one obtains an expression that can be
written
\begin{eqnarray}
U_0&=&\frac{2}{\pi gh}
\bigg\{-\frac{h^2}{4}\int^{R_+}_{R_-}
\frac{R^4 dR}{\sqrt{(R^2_+-R^2)(R^2-R_-^2)}}\nonumber\\
&+&D(D-C)\int^{R_+}_{R_-}\frac{dR}{\sqrt{(R^2_+-R^2)(R^2-R_-^2)}}\nonumber\\
&+&\bigg[1-\frac{h(C+D)}{2}\bigg]
\int^{R_+}_{R_-}\frac{R^2 dR}{\sqrt{(R^2_+-R^2)(R^2-R_-^2)}}\bigg\}.
\label{A.1}
\end{eqnarray}
Using formulas 218.01 and 218.00 of ref.\cite{elliptic}
one can obtain
\begin{equation}
\int^{R_+}_{R_-}\frac{R^2 dR}{\sqrt{(R^2_+-R^2)(R^2-R_-^2)}}=R_+E(k)
\label{A.2}
\end{equation}
and
\begin{equation}
\int^{R_+}_{R_-}\frac{dR}{\sqrt{(R^2_+-R^2)(R^2-R_-^2)}}=\frac{1}{R_+}K(k).
\label{A.3}
\end{equation}
Also using formulas 218.06 and 314.04 of ref.\cite{elliptic} one
finds
\begin{equation}
\int^{R_+}_{R_-}\frac{R^4 dR}{\sqrt{(R^2_+-R^2)(R^2-R_-^2)}}
=\frac{1}{3}R^3_+\bigg[-(1-k^2)K(k)+2(2-k^2)E(k)\bigg].
\label{A.4}
\end{equation}
Inserting these expressions into eq.(\ref{A.1}) one obtains
eq.(\ref{21}).

\end{appendix}

\newpage
\centerline{\bf Figure Captions}
\vspace{0.4cm}
\noindent
{\bf Fig. 1(a)}

\noindent
The period $P$ of the bounce as a function
of the parameter $C$ for $D=1$ and $h=0.7$.

\noindent
{\bf Fig. 1(b)}

\noindent
The sphaleron action and classical action per unit length
as function of temperature $T$
for $D=1$ and $h=0.7$.

\noindent
{\bf Fig. 2(a)}

\noindent
The period $P$ of the bounce as a function
of the parameter $C$ for $D=1$ and $h=0.3$.

\noindent
{\bf Fig. 2(b)}

\noindent
The sphaleron action and classical action
per unit length  as function of temperature $T$
for $D=1$ and $h=0.3$.

\noindent
{\bf Fig. 3(a)}

\noindent
The parameter $C$ as solution of $dP/dC =0$,
$P$ the period of the bounce, plotted as a
function of $h$ for $D=1$. 
The dotted line represents a maximum of $C$ at
given $h$. 
One should observe
the disapppearance of the bifurcation point
at $C=1.2$ (cf. Fig. 2(a)) beyond
$h=0.5$ in agreement with the criterion
of Section 5.

\noindent
{\bf Fig. 3(b)}

\noindent
The parameter $C$ as solution of $dP/dC =0$,
$P$ the period of the bounce, plotted as a
function of $h$ for $D=2$. 
The dotted line represents a maximum of $C$ at 
given $h$.
One should observe
the disapppearance of the bifurcation point beyond
$h=0.25$ in agreement with the criterion
of Section 5.

\noindent
{\bf Fig. 4}

\noindent
The shape of the brane at the bifurcation point 
for $D=1, h=0.3, C=1.20434$.




\newpage
\epsfysize=8cm \epsfbox{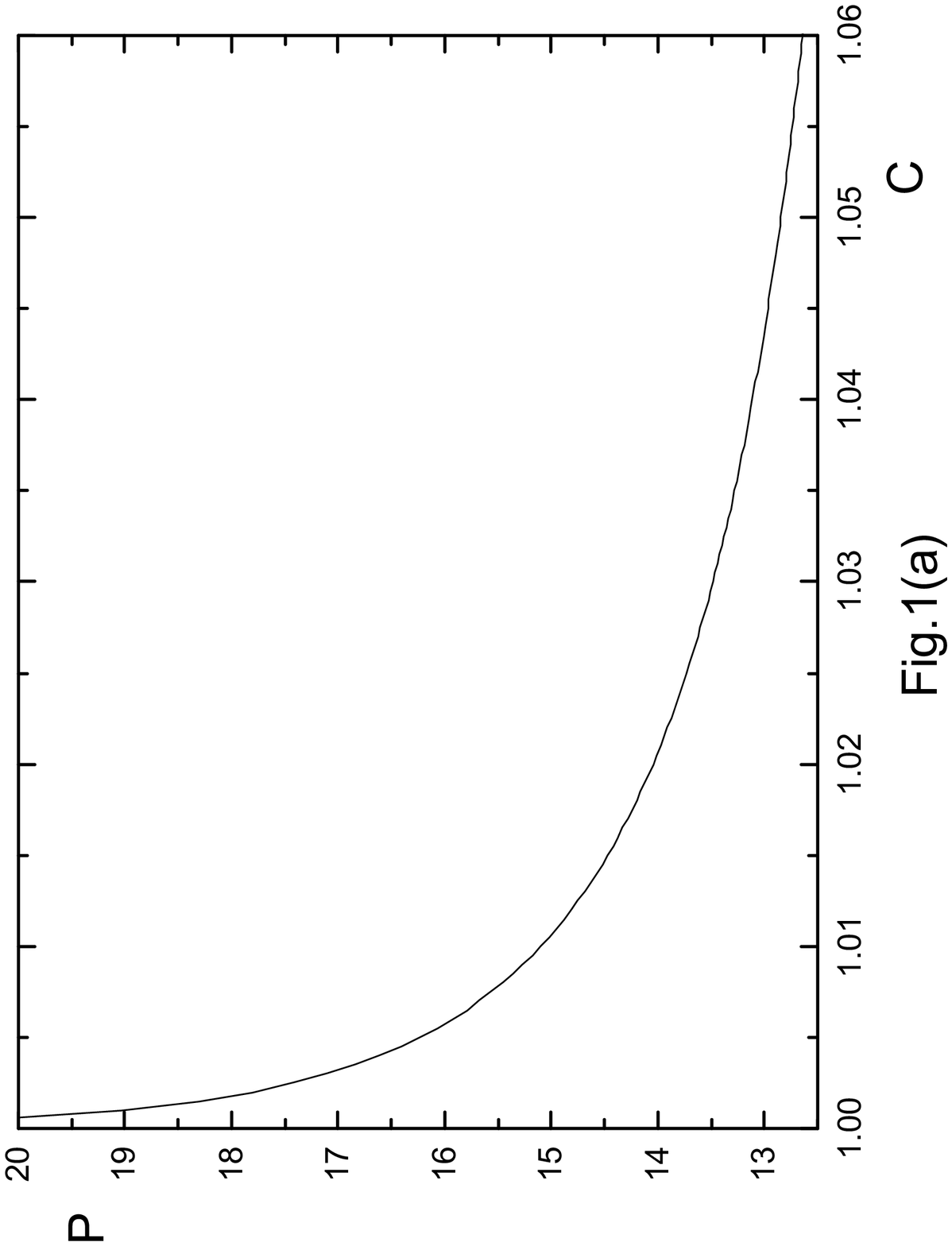}
\newpage
\epsfysize=8cm \epsfbox{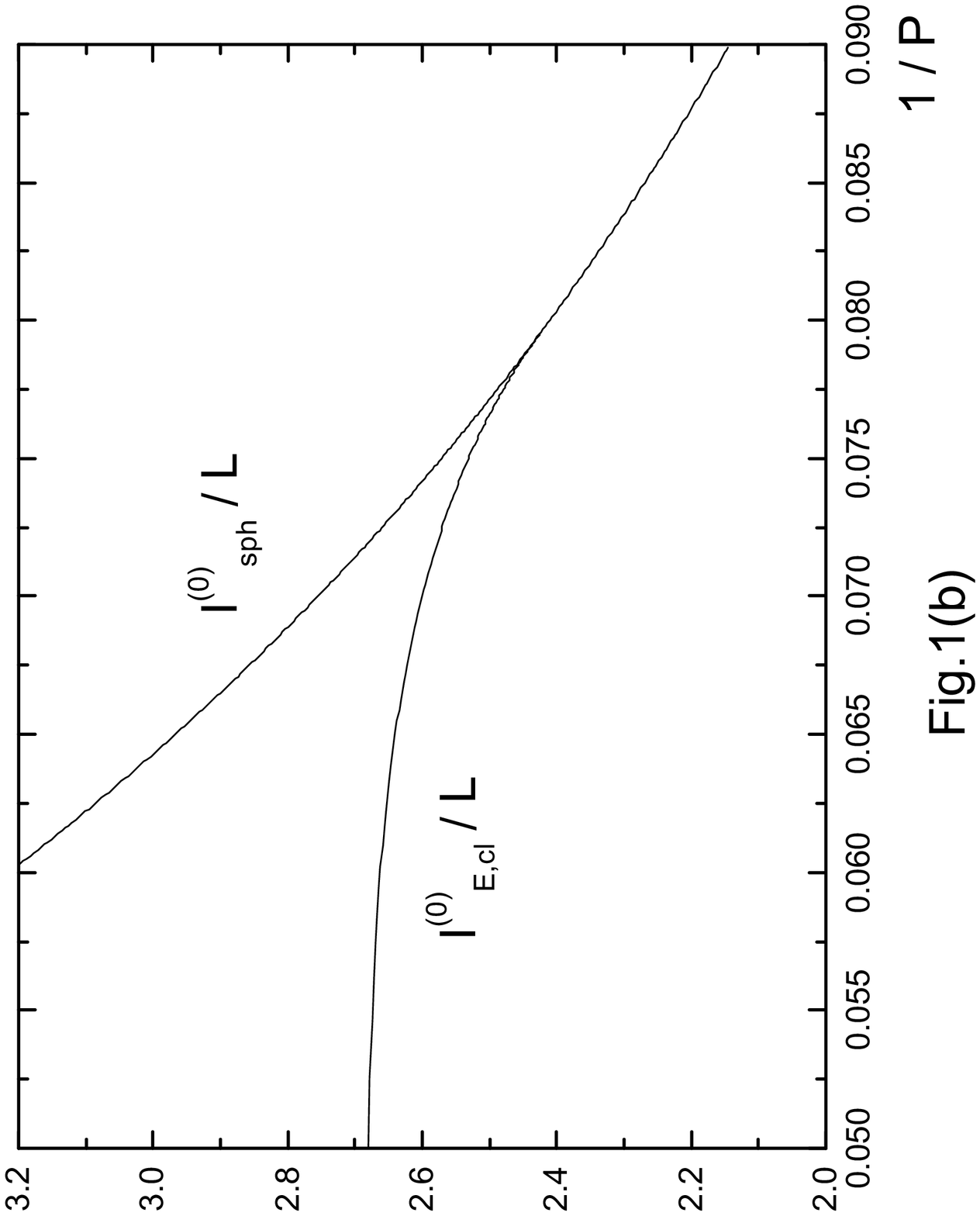}
\newpage
\epsfysize=8cm \epsfbox{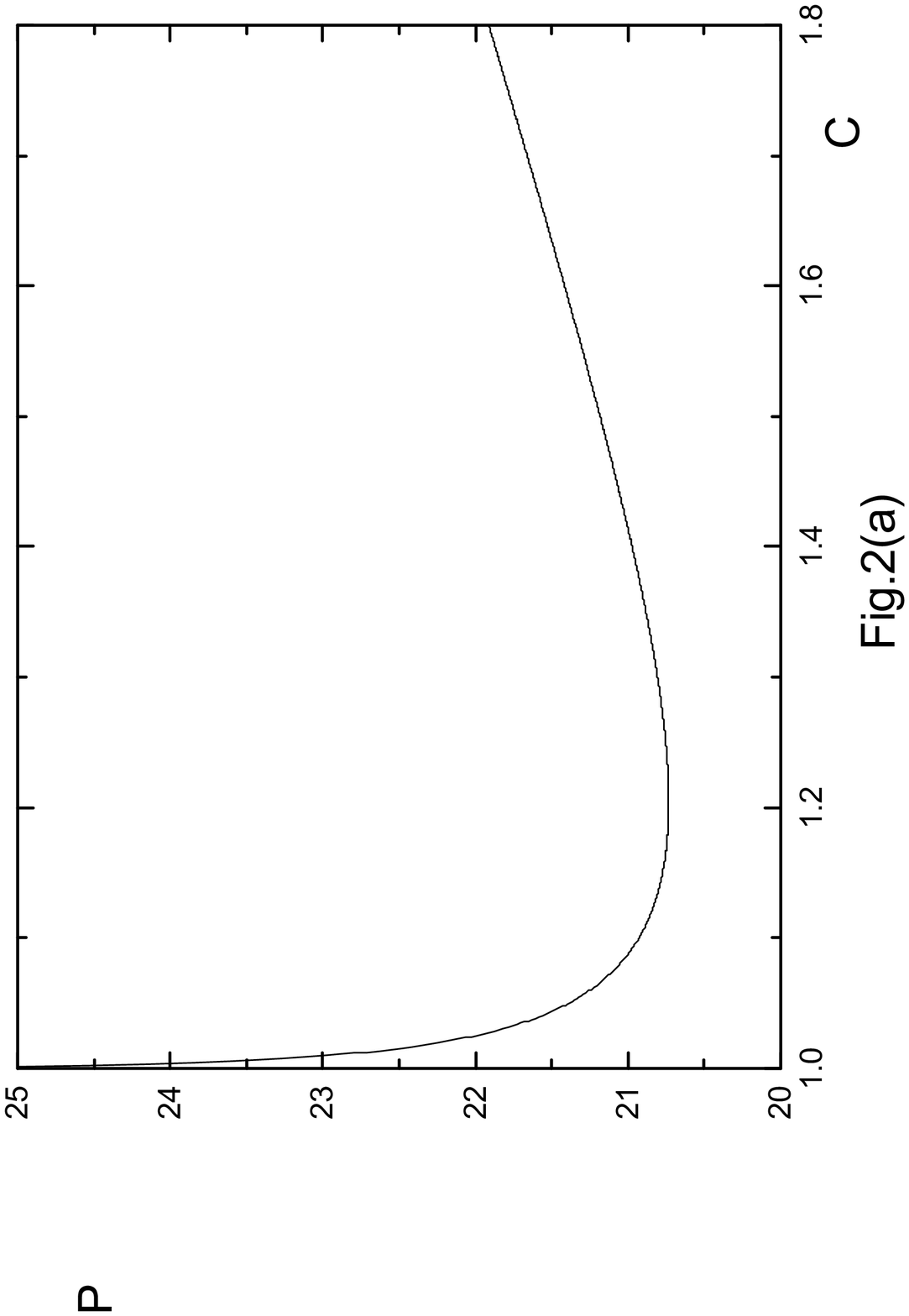}
\newpage
\epsfysize=8cm \epsfbox{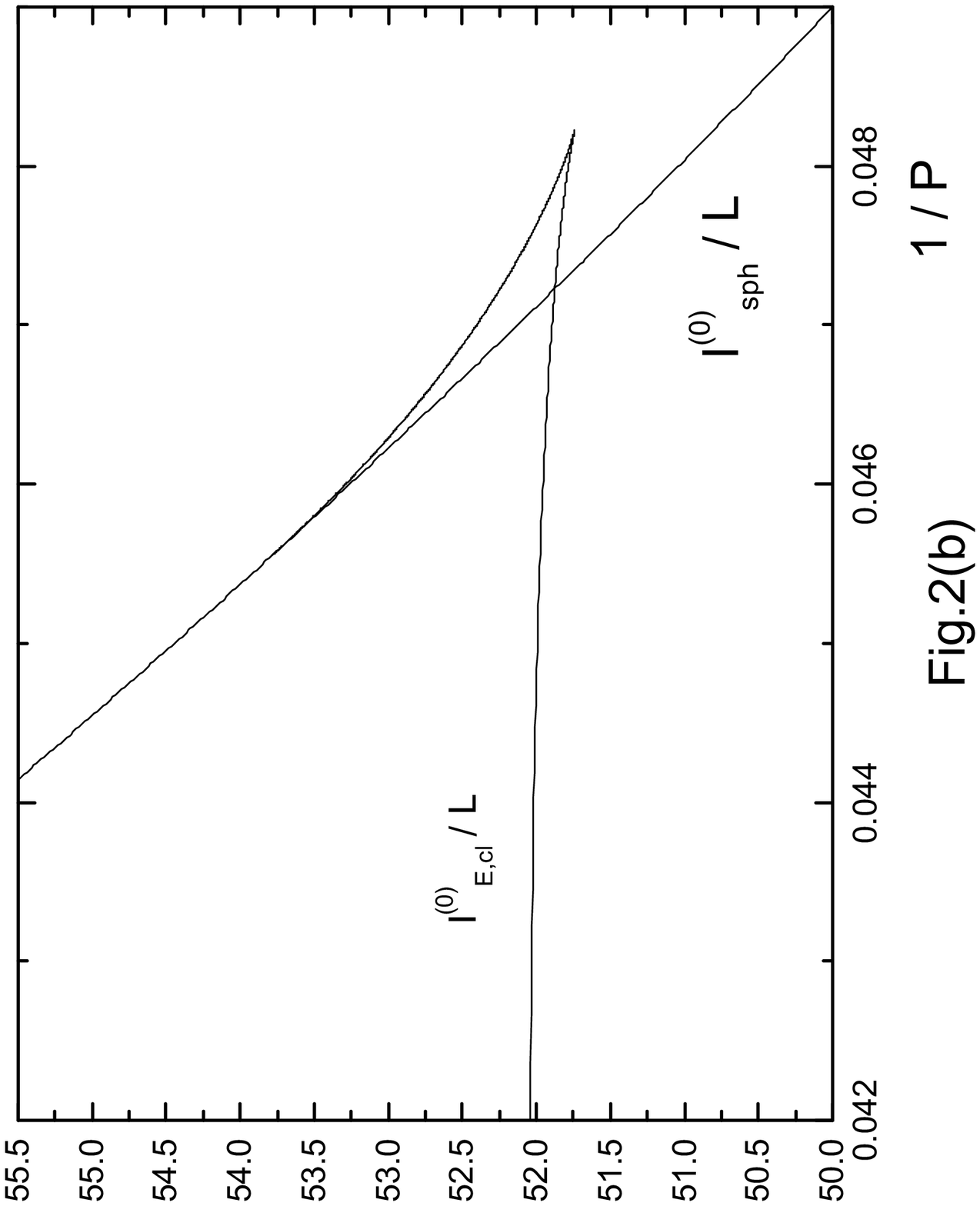}
\newpage
\epsfysize=8cm \epsfbox{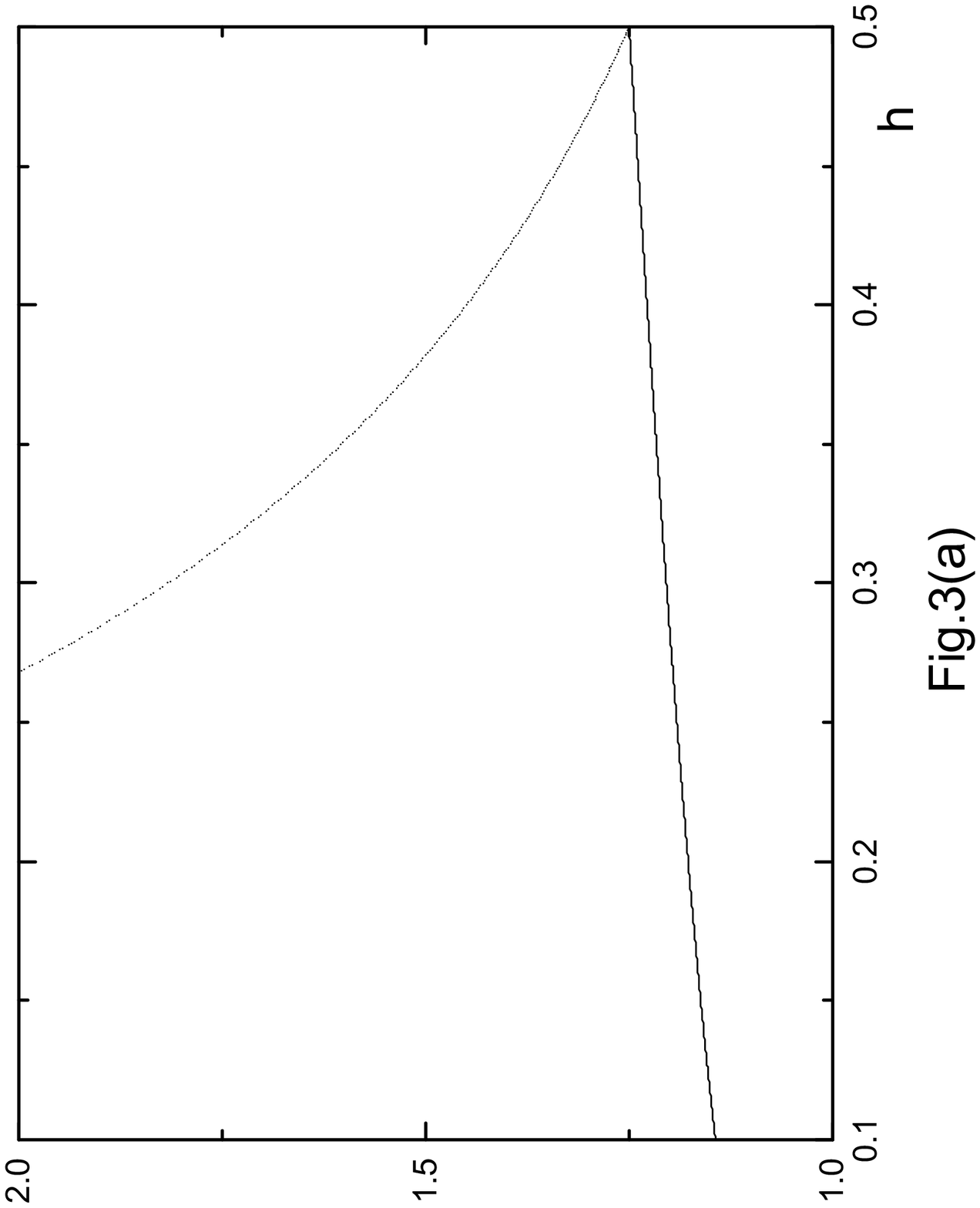}
\newpage
\epsfysize=8cm \epsfbox{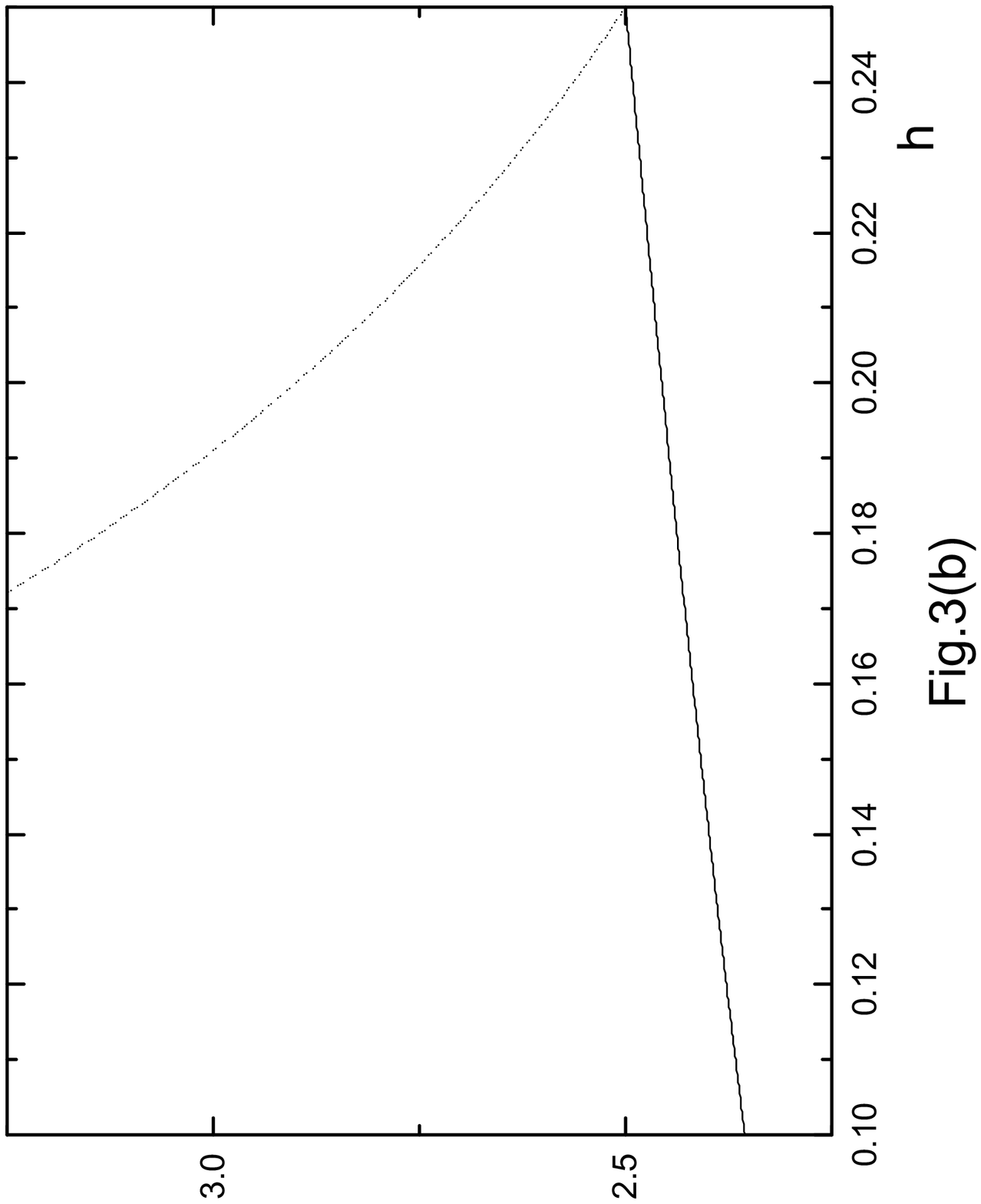}
\newpage
\epsfxsize=15cm \epsfbox{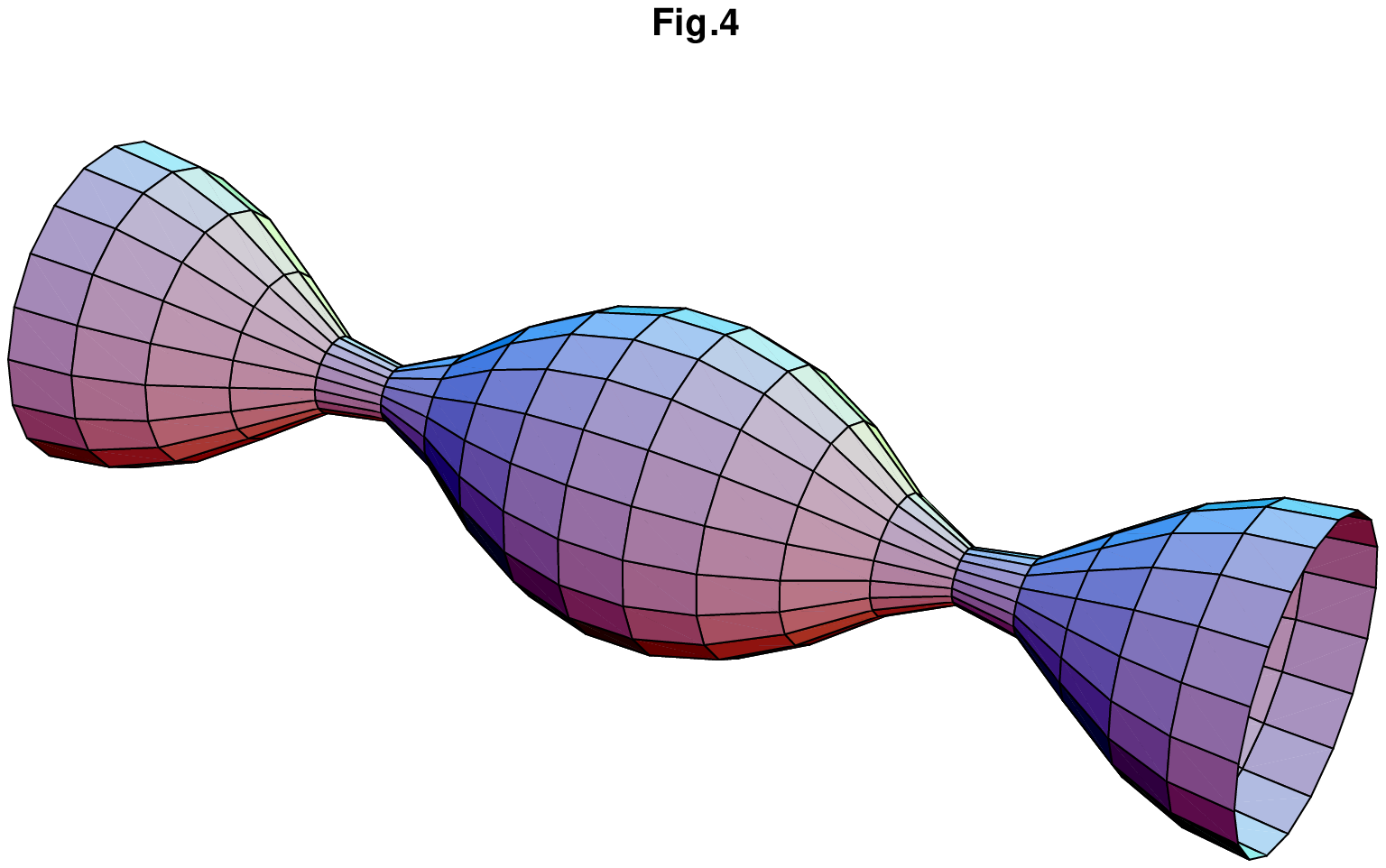}

\end{document}